\documentclass[aps,pre,floats,twocolumn,twoside,floatfix]{revtex4}

\usepackage{graphicx}
\usepackage{amsmath,amssymb}
\usepackage{psfrag}
\usepackage[utf8]{inputenc}

\def\be{\begin{equation}}
\def\ee{\end{equation}}
\def\bea{\begin{eqnarray}}
\def\eea{\end{eqnarray}}

\begin{document}

\title{Microscopic models of mode-coupling theory: the ${\mathsf
    F}_{12}$ scenario}

\author{Jeferson J. Arenzon}

\affiliation{Instituto de F{\'\i}sica, Universidade Federal do Rio
  Grande do Sul \\ CP 15051, 91501-970 Porto Alegre RS, Brazil}
\affiliation{LPTHE, Universit\'e Pierre et Marie Curie-Paris VI, 
4 Place Jussieu, FR-75252 Paris Cedex 05, France}

\author{Mauro Sellitto}

\affiliation{Department of Information Engineering \\ Second
  University of Naples, I-81031 Aversa (CE), Italy}

\newcommand{\tc}{t_{\scriptscriptstyle \rm c}}
\newcommand{\tw}{t_{\scriptstyle \rm w}}
\newcommand{\kB}{k_{\scriptscriptstyle \rm B}}
\newcommand{\Ttric}{T_{\scriptstyle \rm tric}}
\newcommand{\Tc}{T_{\scriptstyle \rm c}}
\newcommand{\Tg}{T_{\scriptstyle \rm c}}
\newcommand{\TK}{T_{\scriptscriptstyle \rm K}}
\newcommand{\pc}{p_{\scriptstyle \rm c}}
\newcommand{\qc}{q_{\scriptstyle \rm c}}
\newcommand{\qEA}{q_{\scriptstyle \rm EA}}
\newcommand{\Phic}{\Phi_{\scriptstyle \rm c}}
\newcommand{\fEA}{f^{\scriptscriptstyle \rm EA}}

\begin{abstract} 
  We provide extended evidence that mode-coupling theory (MCT) of
  supercooled liquids for the ${\mathsf F}_{12}$ schematic model
  admits a microscopic realization based on facilitated spin models
  with tunable facilitation. Depending on the facilitation strength,
  one observes two distinct dynamical glass transition lines--continuous
  and discontinuous--merging at a dynamical tricritical-like point
  with critical decay exponents consistently related by MCT
  predictions. The mechanisms of dynamical arrest can be naturally
  interpreted in geometrical terms: the discontinuous and continuous
  transitions correspond to bootstrap and standard percolation
  processes, in which the incipient spanning cluster of frozen spins
  forms either a compact or a fractal structure, respectively.  Our
  cooperative dynamical facilitation picture of glassy behavior is
  complementary to the one based on disordered systems and can account
  for higher-order singularity scenarios in the absence of a finite
  temperature thermodynamic glass transition.  We briefly comment on
  the relevance of our results to finite spatial dimensions and to the
  ${\mathsf F}_{13}$ schematic model.
\end{abstract}

\maketitle

\section{Introduction}

The glassy state of matter remains an active area of research despite
decades of study.  From a theoretical viewpoint, the most fundamental
issue concerns the very nature of the vitrification process, that is
whether the amorphous state represents a genuine thermodynamic phase
or rather a purely dynamical accident.  It is well know that the issue
is made especially difficult by the lack of a suitable observable
allowing the unambiguous identification of the elusive {\it amorphous
  order}, and the exceedingly long equilibration times involved in
experiments and simulations. The situation is further complicated by
the fact that theoretical modelling of glassy systems has made clear
that slow relaxation phenomena are ubiquitous and may result from very
distinct microscopic mechanisms~\cite{BiKo11}.  Thus, it may appear
rather unlikely that a single theoretical framework encompasses the
large variety of behaviors observed in glassy materials.  For this
reason, one is often focused on the attempt to predict those features
that are thought to be universal in a minimal setting, which is the
basic philosophy we adopt here.

Mode-coupling theory (MCT), is considered by many the most
comprehensive first-principle approach to the dynamics of supercooled
liquids (for reviews, see~\cite{GoSj92,Gotze09,Das04,ReCh05}).
Despite some limitations, it has been much successful in predicting
peculiar higher-order singularities which have been confirmed by
experiments and computer
simulations~\cite{KoAn94,KoAn95,Dawson00,Pham02,EcBa02,ScTa05,Krakoviack07,KiMiSa11,KuCoKa11}.
A key tenet of MCT is that the glass transition is a purely {\em
  dynamical} phenomenon unrelated to any thermodynamic
singularity~\cite{Leutheusser84,BeGoSj84,Kirkpatrick85}. This
prediction is rather puzzling from a statistical mechanics point of
view because diverging relaxation time-scales are typically associated
with diverging lengthscales~\cite{MoSe06}, and the latter are a consequence of a
thermodynamical phase transition, as clearly exemplified by
traditional critical phenomena.  It has been argued that this unusual
situation stems from uncontrolled approximations, which crucially miss
thermally activated hopping processes responsible for low-temperature
equilibration.  Therefore, the dynamical nature of the glass transition
singularity has been much debated since its first
appearance~\cite{Siggia85,Das85b} and much work has been devoted to
clarify the status of
MCT~\cite{KiWo87,KiTh87,KiThWo89,BoCuKuMe96,MePa99b,MaMiRe06,BaBaWo08,AnBiBo09,ScSc11,IkMi11,FrRiRiPa11}.

Fruitful developments in the dynamics of mean-field disordered systems
have subsequently shown the intimate connection of some spin-glass
models with MCT and led to the formulation of the so-called random
first-order transition (RFOT)
approach~\cite{KiWo87,KiTh87,KiThWo89}. The most important ingredient
brought about into the discussion is the existence of an extra {\em
  thermodynamic} glass phase at a Kauzmann temperature $\TK$ below the
dynamical glass transition predicted by MCT.  This has immediately
suggested that while the dynamical transition is an artifact of
mean-field approximation (and therefore doomed to disappear in real
systems), genuine glassy behavior would be a more profound reflection
of an underlying non-trivial Gibbs measure corresponding to a one-step
replica-symmetry breaking spin-glass phase.  Such developments have
been generalized in several ways and have led to a low-temperature
extension of MCT to off-equilibrium situations, revealing new
interesting features such as aging
phenomena~\cite{CuKu93,CuKu94,FrMe94a,FrMe94b} and effective
temperature~\cite{CuKuPe97,Cugliandolo11}.

The thermodynamic RFOT perspective is in apparent conflict with a
different line of thought inspired by facilitated (or kinetically
constrained) models first introduced by Fredrickson and
Andersen~\cite{FrAn84} (see Ref.~\cite{RiSo03} for a review).  These
models have a trivial thermodynamics, and so their glassy behavior has
a purely dynamical origin by construction.  For this reason, they
appear to be at odds with any thermodynamic perspective on the glass
transition problem. Moreover, since the simplest (so-called
non-cooperative) version of such systems does not have neither a
thermodynamic nor a dynamical transition, the facilitation approach
also appears to be inconsistent with MCT.  Therefore, the notion of
dynamical facilitation seems to provide an alternative way to the
glass formation quite in contrast to the MCT and RFOT approach (a
viewpoint advocated, e.g., in Ref.~\cite{GaCh10}).  It has long been
noticed, however, that several peculiar features of MCT and RFOT are
also present in facilitated and kinetically constrained
systems~\cite{KuPeSe97,Sellitto02,SeBiTo05}.  In fact, we have
recently shown that one can construct a class of facilitated spin
systems that actually reproduces the two most relevant MCT
scenarios~\cite{SeMaCaAr10,Sellitto12}.  This means that the dynamic
facilitation approach on one hand, and the MCT and RFOT scenario on
the other, should not be considered as alternative but rather as
complementary representations of glassy dynamics. Our statement is
consistent with recent results showing that, in a specific case, there
is an exact mapping between facilitated and disordered spin
systems~\cite{FoKrZa12}.

In this paper we present detailed numerical results supporting the
idea that cooperative facilitated systems provide a microscopic
realization of MCT including higher-order bifurcation singularity.
The rest of the paper is organized as follows.  To set the stage, in
the section~2, we briefly review the basic predictions of MCT 
approach to glass transition.  In the section~3, we
introduce facilitated spin systems and recall their connections to the
bootstrap percolation problem on a Bethe lattice.  In the section~4 we
define the model we study in this paper and report the exact results
for the phase diagram and the arrested part of correlations.
Numerical simulation for the relaxation time and the critical decay
law exponents are presented and compared with MCT predictions in the
section~5.  Finally, in the section~6 we present the conclusions and
possible directions of future works including a brief discussion on
the possible relevance of our results to finite spatial dimension and
to the ${\mathsf F}_{13}$ schematic model.  Some technical
calculations are reported in the appendix.

\section{Mode-coupling theory: the ${\mathsf F}_{12}$ scenario}

A comprehensive description of MCT can be found in
Refs.~\cite{Gotze09}. Here, we outline only those aspects which are
pertinent to the present work and that are useful for comparison with
our theoretical and simulation results. In particular, since we focus
on schematic models we shall disregard the wavevector dependence of
the relevant physical quantities.
In the MCT, the microscopic system dynamics is projected, through the
Zwanzig-Mori formalism, on a set of observables which is thought to be
physically relevant, namely the local density fluctuations,
$\delta\rho(t)$.  The correlator of these quantity $\phi(t) =
\langle\delta\rho(t)\delta\rho(0)\rangle/\langle|\delta\rho|^2\rangle$
is shown to exactly obey the evolution equation:
\begin{eqnarray}
  \phi(t) + \tau_0 \dot \phi(t) + \int_0^t m(t-s)\ \dot{\phi}(s)\,
  \mathrm{d}s &=& 0,\label{MCT}
\end{eqnarray}
where $\tau_0$ is the characteristic microscopic timescale of the
system and we assume, for simplicity, overdamped local motion
appropriate e.g. to hard-sphere colloids (for molecular liquids one
should add an inertial term).  The memory kernel $m(t-s)$ takes into
account the retarded friction effect which arises from the caging of a
particle by its neighbors. It is controlled by the static structure
factor $S(q) = (1/N)\langle|\delta\rho|^2\rangle$ and does not involve
any thermodynamic singularity.  In the dilute limit the memory
function is negligible and the relaxation is exponential. At
high-density or low-temperature, the memory kernel cause a viscosity
increase through a feedback mechanism that eventually leads to the
structural arrest of fluctuations.  For a general system, the explicit
expression of $m(t-s)$ is rather complicated, and in order to obtain
quantitative
predictions, one is forced to introduce some approximations.  In
schematic models the memory function is approximated by a low order
polynomial of correlators, with coupling constants that depend solely
on $S(q)$.
A key quantity in describing the liquid-glass transition is the
so-called nonergodicity parameter (also known as Edwards-Anderson
parameter in the spin-glass literature), that is the long-time limit
of the correlator,
\begin{eqnarray}
\Phi &=& \lim_{t \to \infty } \phi(t).
\end{eqnarray}
In the fluid state, the system is ergodic and $\phi(t)$ decays to zero
with time, $\Phi=0$.  When $\Phi$ is finite the system is unable to
fully relax, its dynamical is arrested and the liquid becomes a glass.
For schematic models one can easily derive the liquid-glass phase
diagram. Taking the long-time limit of Eq.~(\ref{MCT}), one obtains
the bifurcation equation:
\begin{eqnarray}
  \frac{\Phi}{1-\Phi} &=& m\left[\Phi\right] ,
\end{eqnarray}
whose solutions possibly give different types of ergodic-nonergodic
transitions.  Of interest here is the so-called ${\mathsf F}_{12}$ schematic
model which is defined by the memory function
\begin{eqnarray}
  m \left[ \phi(t) \right] & = & v_1 \phi(t) + v_2 \phi^2(t),
  \label{F12}
\end{eqnarray} 
where $v_1,\, v_2$ are the coupling parameters controlling the system
state. In this case, by solving the bifurcation equation one finds a
{\it discontinuous} liquid-glass transition line, $v_1^c = \sqrt{4
  v_2}-v_2$ with $v_2 \in [1,4]$, at which $\Phi$ jumps from zero to a
finite value $\Phic$, with a square-root singularity, $\Phi -\Phic
\sim \epsilon^{1/2}$, where $\epsilon$ is the distance from the
transition line. For $v_2 \in [0,1]$ one finds a {\it continuous}
liquid-glass transition, $v_1^c=1$, across which $\Phi$ smoothly
departs from zero in a linear fashion, $\Phi \sim \epsilon$.

On the liquid side of the relaxation dynamics, MCT makes several
specific predictions. Near the discontinuous transition line, the
relaxation has a peculiar two-step form. The approach and departure
from the critical plateau at $\Phic$, respectively defining the
$\beta$ and $\alpha$ relaxation regimes, are described by the
so-called critical decay laws:
\begin{equation}
\left| \phi(t)-\Phic \right|\sim  \left\{
\begin{array}{l}
  \displaystyle (t/\tau_{\scriptscriptstyle\beta})^{-a}, \qquad
   \beta\mbox{\rm -regime}; \\ \displaystyle (t/\tau)^{-b}, \,\,\,
    \qquad \alpha \mbox{\rm -regime} .
\end{array}
\right.
\label{eq.alphabeta_regimes}
\end{equation}
The two characteristic times $\tau$ and $\tau_{\scriptscriptstyle
  \beta}$, respectively associated with the $\alpha$ and $\beta$
regimes, increase as power-laws near the critical line:
\begin{equation}
  \tau \sim \epsilon^{-\gamma}, \qquad \tau_{\scriptscriptstyle\beta}
  \sim \epsilon^{-1/2a}.
  \label{eq.tau_beta}
\end{equation}
The exponents $a$, $b$ and $\gamma$ are not independent but obey the
relations:
\begin{equation} 
  \lambda = \frac{\Gamma^2(1-a)}{\Gamma(1-2a)} =
  \frac{\Gamma^2(1+b)}{\Gamma(1+2b)}, \qquad \gamma =
  \frac{1}{2a}+\frac{1}{2b},
  \label{eq.gamma}
\end{equation} 
where $\Gamma$ is the Euler's gamma function.  The correlator presents
universal scaling (the so-called time-temperature superposition
principle) in the late $\alpha$-regime: rescaling time by the
structural relaxation time $\tau$, one obtains that data for
correlator near the critical line can be collapsed onto a single
master function.

Near the continuous transition line there is no $\alpha$-regime and
one finds a single step relaxation with a weak long time tail
described by the critical decay law of the $\beta$-regime, i.e. the
first of Eq.~(\ref{eq.alphabeta_regimes}) with $\Phic =0$ remains
valid. However, the exponent characterizing the relaxation time is
twice that of the discontinuous transition:
\begin{eqnarray}
  \tau_{\scriptscriptstyle\beta} \sim \epsilon^{-1/a}.
\end{eqnarray}
It is important to notice that the exponent $a$ (and therefore $b$ and
$\gamma$) depends on the actual location of the critical point along the
glass transition lines. So this provide us with the possibility of
carefully checking the peculiar MCT prediction connecting the
exponents of the critical decay law and the relaxation time near the
transition over a wide range of parameters.  In particular, when the
point $(v_1^*,v_2^*)=(1,1)$ separating the discontinuous and
continuous transition is approached, $a$ tends to vanish. Near this
glass singularity, which is also called a degenerate A$_3$ point, the
power-law behavior turns into logarithmic relaxation.

\section{Facilitated spin systems and bootstrap percolation on Bethe lattice}

The basic assumption of the dynamical facilitation approach is that, on
a suitable coarse-grained lengthscale, one can model the structure of
a liquid by an assembly of higher/lower density mesoscopic cells with {\em
  no} energetic interaction. A binary spin variable
is assigned to every cell depending on its solid- or
liquid-like structure.  The next crucial step is to postulate that
there exists a timescale over which the effective microscopic dynamics
can be mapped onto a `simple' form: local changes in cells structure
occur if and only if there is a sufficiently large number, say $f$, of
nearby liquid-like cells ($f$ is called the facilitation parameter).  The
latter assumption is admittedly quite remote from the actual liquid
dynamics, and indeed very difficult to derive by analytical
means. Nevertheless, it can be justified on physical grounds: it
mimics the cage effect and gives rise to a large variety of
remarkable, and sometimes unexpected, glassy features, even if the
thermodynamics is completely trivial.  They include stretched
exponential relaxation, super-Arrhenius equilibration time, physical
aging, effective temperature, exponential multitude of blocked states,
dynamical heterogeneity, etc.

Facilitated spin models consist of $N$ non-interacting spins
$\sigma_i=\pm 1$, $i=1,\dots,N$ with Hamiltonian 
\begin{equation}
  {\cal H} = - \frac{h}{2} \sum_{i=1}^N \sigma_i.
\end{equation}
Spins evolve according to a Metropolis-like dynamics: at each time
step a randomly chosen spin is flipped with transition probability:
\begin{eqnarray}
  w(\sigma_i \to -\sigma_i) &=& {\rm min} \left\{ 1, {\rm e}^{- h
    \sigma_i/\kB T} \right\},
\end{eqnarray}
if and only if at least $f$ of its $z$ neighboring spins are in the
state $-1$. When the temperature $T$ is low enough, the fraction of
$-1$ spins is exponentially small, ${\rm e}^{-h/\kB T}$, therefore
spin flipping is rare and sluggish relaxation ensues.  We consider a
Bethe lattice with fixed connectivity, specifically, a regular
$k$-graph with coordination number $z=k+1$: besides one branch going
down, each node has further $k$ branches going up.  The tree-like,
local structure of the Bethe lattice allows for some exact
calculations concerning the asymptotic dynamics.  Let us call $B$ the
probability that, without taking advantage of the configuration on the
bottom, the spin $\sigma_i$ is in, or can be brought to, the state -1 by
only rearranging the sites above it. $B$ verifies the fixed-point
equation:
\begin{eqnarray} 
  B &=& (1-p) + p \sum_{i=0}^{k-f} {k \choose i} B^{k-i} (1-B)^{i}
\label{B} 
\end{eqnarray} 
where $p$ is the probability that the spin is $+1$ in thermal
equilibrium, i.e.:
\begin{eqnarray}
  p &=& \frac{1}{ 1 + {\rm e}^{-1/T} },
\end{eqnarray}
where we set $h/\kB=1$ hereafter.
Eq.~(\ref{B}) is closely related to that of bootstrap percolation
(BP)~\cite{ChLeRe79,SeBiTo05}, a problem that is known to emerge in a
wide variety of contexts (for a review, see
Refs.~\cite{AdLe03,GrLaDa09}).  In BP every lattice site is first
occupied by a particle at random with probability $p$.  Then, one
randomly removes particles which have less than $m$
neighbors.  Iterating this procedure leads to two possible asymptotic
results~\cite{ChLeRe79}. If the initial particle density is larger than
a threshold $\pc$ there is an infinite cluster of particles that
survives the culling procedure, whereas for $p<\pc$ the density of
residual particles is zero.  In Ref.~\cite{ChLeRe79} it has been shown
that the probability $1-B$ that a particle is blocked because it has
at least $m$
neighboring particles above it which are blocked (without taking
advantage of vacancies below) satisfies the self-consistent
Eq.~(\ref{B}).
By exploiting the relation with the BP one can characterize the
dynamics on a Bethe lattice. There are three qualitatively
distinct cases that we now discuss. \\
\subsection{Noncooperative dynamics}

For $f=1$, any spin only needs one nearby down-spin to flip and
the dynamics is called {\it noncooperative}. Eq.~(\ref{B}) becomes
\begin{eqnarray}
  1-B &=& p (1-B),
\end{eqnarray}
obviously implying that there is only the solution $B=1$ for any $p$,
i.e. with probability one each spin can be brought into the up-state
through a finite number of allowed spin-flips.  One can see
that relaxation dynamics has a single-step form with an Arrhenius
equilibration time at any temperature. Thus, the noncooperative case
may be useful 
for modelling {\em strong} glasses.  For
this reason it will not be considered here.  For more complex forms of
glassy dynamics we need local relaxation events involving two or more
facilitating (down) spins, i.e. a cooperative dynamics.

\subsection{Cooperative dynamics}

\subsubsection{Continuous transition} 

For $f=k$ or $k+1$ there is an additional solution, with $B<1$, of
Eq.~(\ref{B}) which is obtained when $p$ is large enough.  The
transition to this regime is {\em continuous}, because for arbitrary
small $1-B$, Eq.~(\ref{B}) becomes
\begin{equation}
    1-B = p \ {k \choose f-1} (1-B)^{k-f+1}+ \dots, 
\label{1-Bsmall}
\end{equation}
which implies $\pc=1/k$ for either $f=k$ or $k+1$. Notice that in
the later case, the equation has the exact same form as for $f=k$.  
In fact, these cases
are completely equivalent to conventional percolation~\cite{ChLeRe79},
and for this reason, near $\pc$ the incipient cluster of permanently
frozen spins (i.e. spins unable to flip because surrounded by more
than $f$ neighboring up spins) has a {\em fractal} structure.  The
facilitated spin dynamics of these systems has never been explored to
our knowledge. We notice that the critical temperature in this case is
negative $1/\Tc=-\ln(k-1)$. This means that in order to study
relaxation dynamics one must start with initial configurations having
a majority of down spins.

\subsubsection{Discontinuous transition}

For $1<f<k$ the power of $1-B$ on the right and left hand side of
Eq.~(\ref{1-Bsmall}) are different and therefore only a {\em
  discontinuous} transition --at a critical value $\pc$ that depends
on $k$ and $f$-- is possible.  This transition is characterized by a
square root singularity coming from low temperatures and is rather
peculiar as it produces an unusual {\em mixed} or {\em hybrid}
behavior: the fraction of frozen spins jumps from zero to a finite
value, with divergent fluctuations as in critical phenomena. This is
much similar to the behavior of the non-ergodicity parameter in MCT
and this form of  dynamical arrest corresponds to the sudden
emergence of a giant cluster of frozen spins with {\em compact}
structure.  The geometric origin of this behavior has been understood
quite in detail as being related to the divergence of the size of {\em
  corona} clusters near the transition~\cite{DoGoMe06,ScLiCh06}.
Several results~\cite{Sellitto02}, including those related to large scale
cooperative rearrangements responsible for slow
dynamics~\cite{MoSe05}, and the TAP-like organization of blocked
states below the threshold~\cite{SeBiTo05}, have already suggested a
close analogy with MCT and mean-field disordered systems.

\section{Facilitated spin mixtures with tunable facilitation: exact results}

The systems considered so far have a fixed facilitation determined by
the integer number $f$, and so they can exhibit either a discontinuous
or a continuous glass transition. In this section we consider a
recently introduced class of facilitated spin systems in which the
facilitation strength can be smoothly tuned~\cite{SeMaCaAr10}. This is
obtained by making the facilitation parameter a lattice site dependent
random variable, in close analogy with the random bootstrap
percolation problem studied by Branco~\cite{Branco93} (see also
Refs.~\cite{BaDoGoMe10,CeLaDaGl11} for recent developments in the
context of complex networks).  This means that for every site $i$ the
facilitation of the spin $\sigma_i$ is a random variable generally
described by the probability distribution
\begin{eqnarray}
  {\cal P}(f_i) & = & \sum_{\ell=0}^{k+1} \ w_{\ell} \ \delta_{f_i,\ell} ,
\label{eq.distf.general}
\end{eqnarray}
where the weights $\{w_{\ell}\}$ controlling the facilitation strength 
satisfy the condition $0 \le w_{\ell} \le 1$ and the normalization
\begin{eqnarray}
\sum_{\ell=0}^{k+1} w_{\ell} = 1.
\end{eqnarray}
By suitably changing the coefficients $\{w_{\ell}\}$ one can thus
explore a variety of different situations, e.g. the
robustness
of the glass phase against dilution and, more interestingly, the
crossover between the discontinuous and continuous glass transition or
the existence of multiple glassy states.  We remark that mixtures of
particles with different kinetic constraints were first considered in
a granular matter context, in the attempt to understand particle
segregation phenomena under gravity by a purely dynamical 
approach~\cite{SeAr00,LeArSe01,FeArLeSe03}).  Hereafter, we shall
denote with $\left\langle \cdots \right\rangle_f$ the average over the
probability distribution Eq.~(\ref{eq.distf.general}).  Following
Refs.~\cite{ChLeRe79,Branco93,SeBiTo05}, we can easily generalize the
results presented in the previous section and evaluate the probability
$B$ as:
\begin{eqnarray}
  B &=& 1-p + p\left\langle \sum_{n=0}^{k-f} \binom{k}{n} B^{k-n}(1-B)^n
  \right\rangle_{\!\!f}.
\end{eqnarray}
As we have already anticipated, the precise nature of the glass
transition depends on the behavior of the fraction of permanently
frozen spins that, in this framework, can be exactly computed from $B$
as follows
\begin{widetext}
  \begin{eqnarray}
    \Phi &=& \left\langle p \sum_{n=0}^{f-1} \binom{z}{n} B^n
    (1-B)^{z-n} + (1-p) \sum_{n=0}^{f-1} \binom{z}{n} (1-h)^n
    h^{z-n} \right\rangle_{\!\!f}, \label{eq.Phi}
  \end{eqnarray}
\end{widetext}
where
\begin{eqnarray}
h &=& p \left\langle \sum_{m=0}^{f-2} \binom{z-1}{m} B^m (1-B)^{z-1-m}
\right\rangle_{\!\!f} .
\end{eqnarray}
The two contributions in Eq.~(\ref{eq.Phi}) represent the probability
that a spin is frozen in the $\pm 1$ state, respectively.
In order to evaluate $\Phi$, in fact, one must take into account
that in the neighborhood of both up and down spin states, there must
be less then $f$ facilitated spins. Let us consider, for clarity, the
homogeneous facilitated case (for heterogeneous facilitation, one just
needs to further average over the distribution Eq.~(\ref{eq.distf.general})).
To be blocked, an up spin (found with probability $p$) may have, among
its $z$ neighbors, up to $f-1$ down spins. These neighbors already
have one spin up (the main spin) and the probability that each one is
facilitated is precisely $B$. The overall contribution of all
permutations is thus given by the first term in the above equation.
When the main spin is down (with probability $1-p$), we write a term
similar to the first one, but replacing $B$ with $1-h$. $h$ is the
probability of a neighbor of the main spin (that is down) being up
(with probability $p$) and blocked (because beyond the main spin, we
consider the possibility of only $f-2$ further down spins).  Notice
that now to evaluate $h$ we consider the possible permutations of $f-2$
down spins among the $z-1$ neighbors.
When the facilitation is no longer homogeneous, we should average both
$\Phi$ and $h$ over the distribution ${\cal P}(f_i)$.

\begin{figure}[htbp]
\includegraphics[width=8.5cm]{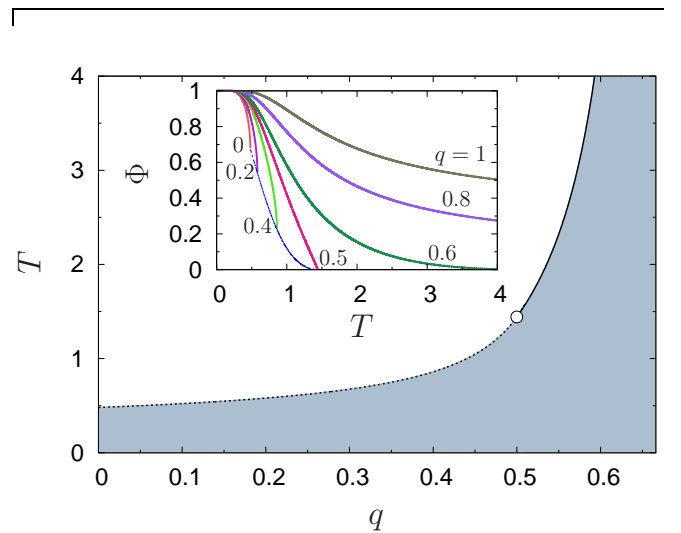}
\caption{Phase diagram for $z=4$ and facilitation as in
  Eq.~(\ref{eq.distf}). The dark region is the glassy phase. The
  dotted/solid line is the hybrid/continuous transition.  Inset:
  Fraction of frozen spins, Eq.~(\ref{eq.ap.Phi234}), vs temperature for
  several values of $q$ and $r=10^{-3}$. Below $q=1/2$, $\Phi$ jumps
  to a finite value at the transition which is represented by the
  dotted line.}
\label{fig.diag}
\end{figure}

The calculation of the phase diagram and the fraction of frozen spins
for different types of mixtures on a Bethe lattice with $z=4$ is
detailed in the appendix. Here, we shall focus on a ternary mixture
with facilitations $f_i$ chosen from the probability distribution
\begin{equation}
  {\cal P}(f_i) = (1-q) \ \delta_{f_i,k-1} + (q-r) \ \delta_{f_i,k}+ r
  \ \delta_{f_i,k+1} ,
\label{eq.distf}
\end{equation}
with $0\leq r\leq q\leq 1$.  The phase diagram for a Bethe lattice
with $z=4$ is depicted in Fig.~\ref{fig.diag}, and comprises two glass
transition lines, of discontinuous and continuous nature, given
respectively by
\begin{equation}
  \frac{1}{\Tg(q)} = \left\{
\begin{array}{l}
  \displaystyle \ln\left( 8-12q \right), 
  \qquad  0 \leq q \leq \frac{1}{2}; \\ \\
  \displaystyle \ln \frac{1}{3q-1}, \,\,\,\,\,\,
  \qquad   \frac{1}{2} \leq q < \frac{2}{3} . 
\end{array}
\right. 
\label{eq.pc}
\end{equation}
The two lines join smoothly at $q=1/2$, corresponding to the dynamical
tricritical temperature
\begin{eqnarray}
  \Ttric & = & 1/\ln 2. 
  \label{eq.trc}
\end{eqnarray}
The critical temperature depends only on the total fraction $q$ of
$f=z-1$ and $z$ overconstrained spins, but not on their relative
importance $r$.
As expected, the glass transition temperature increases with $q$
since the constraints become stronger and the spins more easily
blocked when compared to the pure case $q=0$.
Notice that for $q=2/3$ the critical temperature diverges signalling that
there always exists a finite fraction of permanently frozen spins at
any positive temperature. Of course, the above phase diagram can be
formally extended to negative temperature, i.e. for $2/3< q\leq 1$, by
contemplating the possibility of having an equilibrium distribution of
$-1$ spins larger than 1/2. The highest temperature is then $ -\infty $
and corresponds to a fully ordered spins configuration in the -1
state. This is the initial configuration one has to start with to
properly perform simulations of equilibrium dynamics at negative
temperature. Such a situation will not be considered here.
The precise nature of the two glass transitions in Eq.~(\ref{eq.pc})
depends on the behavior of the fraction of permanently frozen spins
$\Phi$ near $\Tg$ which, in the specific case $z=4$, is given in the
appendix~\ref{ap234}.
The behavior of $\Phi$ as a function of the temperature is shown in
the inset of Fig.~\ref{fig.diag}, for several values of $q$ and with
$r=10^{-3}$. The height of the critical plateau, shown as a dotted
line in the figure, decreases as $q$ increases, going to 0 as $q\to
1/2$.
We remark that the fraction of frozen spins generally depends on $r$
while the phase diagram does not.  For $ 0 \leq q \leq 1/2$, $\Phi$
jumps to a finite value $\Phic=\Phi(\Tg)$ on the transition line,
meaning that the infinite cluster of frozen spins has a compact
structure. This structure turns out to be quite resilient against
random inhomogeneities in the facilitation strength, in a rather large
range of value of $q$.
The critical exponent $\beta$ associated to the order parameter $\Phi$
is obtained by expanding the above equations in the small parameter
$\epsilon=T-\Tg$.
As expected, when $\Phi_c$ has a finite jump, we find $\Phi-\Phic \sim
\epsilon^{\beta}$, with $\beta=1/2$, the typical square-root
dependence well known in MCT and in other systems with a hybrid
transition. 
For $1/2 < q < 2/3$ the glass transition changes nature as $\Phi$
departs smoothly from zero with a power-law behavior $\Phi \sim
\epsilon^{\beta}$ and the correspondence with standard percolation
implies that the giant cluster of frozen spins has a fractal
structure. Interestingly, in this case $\beta$ depends on $r$: for
$r=0$ one has $\beta=2$, while as soon as a negligibly small amount of
spins with $f_i=4$ is introduced into the system, $r>0$, one has
$\beta=1$.  The robustness of the latter behavior reflects the fact
that the mass of the fractal cluster of frozen spins is essentially
dominated by the {\em dangling ends}, that is, those parts of the
cluster which are connected to the backbone by a single frozen
spin.  Only when $r=0$, dangling ends are completely removed from the
infinite cluster of frozen spins, and the exponent changes to
$\beta=2$~\cite{Branco93}.
Hence, the general scenario emerging for an arbitrary ternary mixture
is that of two distinct glass transitions with $\beta=1/2$ (for the
discontinuous case), and $\beta=1$ (for the continuous one).  These
critical exponents reproduce exactly the MCT results for the ${\mathsf F}_{12}$
schematic model~\cite{Gotze09}. 
Clearly, it is important to establish how these exponents are attained
when the critical line is approached, Also, it is by no means obvious
that systems sharing the same order parameter exponents in the glassy
phase will also have identical critical dynamical exponents.  This type
of problems will be addressed in the next section.

\section{Numerical simulations of a spin mixture with 
  discontinuous and continuous glass transitions}

Since the models we have introduced in the previous section are only
partially solvable, we now turn to numerical simulation. This will
give us the possibility of verifying the power-law behavior of
relaxation time with temperature and of critical decay with time and,
most importantly, the peculiar prediction of MCT that connects the
exponents of these two laws near the continuous and discontinuous
transition line.  One obvious advantage of working with facilitated
spin models is that one can immediately overcome the problem of the
long simulation times needed for thermal equilibration. Moreover, the
exact knowledge of both the critical temperature and the arrested part
of correlations makes the procedure for estimating power-law exponents
much more reliable, as it decreases the number of fitting parameters
from three to one. In fact, in real systems, for a given time window,
the fitting parameters are correlated: a change of the critical point
position in the fit can be compensated, within certain limits, by a
change in the power-law exponent. For this reason, one hardly find
data, which test the critical decay for more than a two decade
time-window.  It is also important to remark that testing MCT in
systems which are described by a somewhat artificial kinetic rule is
particularly interesting because it allows us to probe the degree of
universality of MCT prediction beyond the original context (the actual
Newtonian or Brownian liquid dynamics) and the closure approximations
in which they were originally derived.

\subsection{Persistence}

To explore those universal features of relaxation which are relevant
for a comparison with MCT we focus on the time evolution of
persistence $\phi(t)$, i.e. the probability that a spin has never
flipped between times $0$ and $t$. It is a convenient and widely
used characterization of the dynamics of facilitated spin systems.
The long-time limit of $\phi(t)$ plays the role of the nonergodicity
parameter and represents 
the fraction of permanently frozen spins
\begin{eqnarray}
\Phi = \lim_{t \to \infty} \phi(t).
\end{eqnarray}
When $\Phi=0$ the system is a liquid, while for $\Phi>0 $ ergodicity
is broken and the system is a glass.
\begin{figure}[hbtp]
  \includegraphics[width=8.5cm]{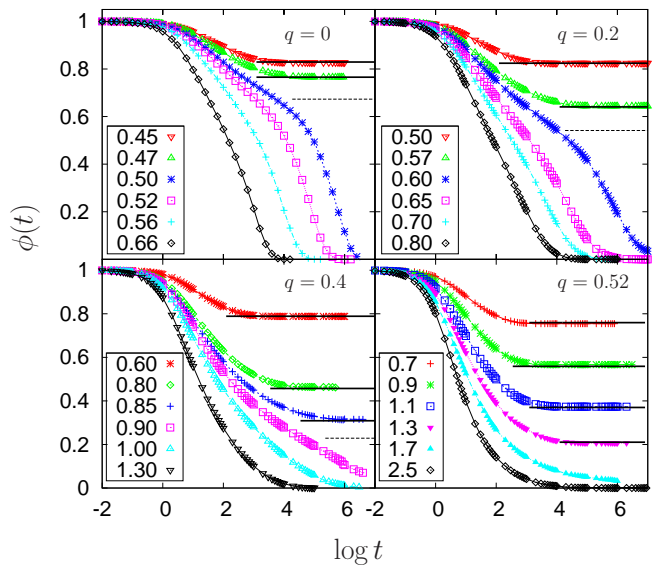}
  \caption{Persistence vs time for several values of $q$ both in the
    discontinuous ($q=0$, 0.2 and 0.4) and the continuous region
    (0.52) of the phase diagram, Fig.~\ref{fig.diag}, for $r=10^{-3}$
    and several temperatures, either above and below $\Tg(q)$. The
    solid black lines show the theoretical prediction,
    Eq.~(\ref{eq.ap.Phi234}), for the plateau heights for $T<\Tg(q)$
    while the dotted ones signal the critical plateau at $T=\Tg(q)$,
    $q<1/2$. For temperatures above the transition, the long time
    value of the persistence is zero.}
  \label{fig.pers_f2f3f4}
\end{figure}

We have been careful enough to disallow initial system configurations
with a fraction of permanently frozen spins that would induce
dynamical reducibility problems in our simulations (see the discussion
in Ref.~\cite{RiSo03} for details concerning these issues). To this
end we have chosen a ternary mixture with a rather negligibly small
fraction of spins with $f=4$, typically $r=10^{-3}$, and not too large
values of $q$ (that is a fraction of spins with $f=3$).  In fact, we
observed that the system becomes too overconstrained and reducibility
effects begin to appear above $q \simeq 0.55$, at least for the system
sizes, $N=10^5-5 \times 10^5$, we have explored in our numerical
simulations.  At each time unity, called a Monte Carlo step (MCS), $N$
attempts to flip an spin are performed.
Simulation results for the persistence $\phi(t)$ are shown in
Fig.~\ref{fig.pers_f2f3f4} for various values of $q$, both above and
below the glass transition line $\Tg(q)$. 
We see that above $\Tg(q)$, every spin is able to flip so that the
long-time limit of persistence is zero, confirming theoretical
expectations and the absence of dynamical reducibility effects. Whereas
for temperatures inside the glassy phase, $T<\Tg(q)$, ergodicity is
broken and the persistence attains a finite plateau which is in
excellent agreement with the value of $\Phi$ analytically computed in
Eq.~(\ref{eq.ap.Phi234}).
We also observe that the height of the critical plateau $\Phic$ (shown
as a dotted line in the figure), which is finite for $q<1/2$, moves
towards zero as $q$ approaches 1/2, and remains zero above it. In the
previous section we have calculated how this $\Phic$ is approached
when $T\to\Tg(q)$ and we now compare the exact results with the
numerical observations.

\begin{figure}[htbp]
\includegraphics[width=8.5cm]{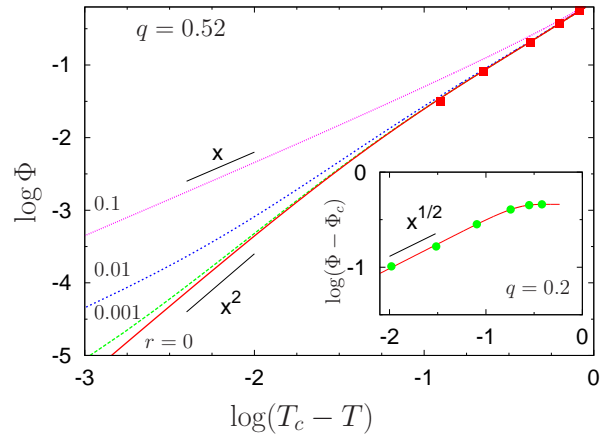}
\caption{Asymptotic fraction of blocked spins $\Phi$ (plateau height)
  as the temperature approaches $\Tg$ for several values of $r$ and
  $q=0.52$ in the continuous region of the phase diagram. Points are
  the result of simulations while the lines are from
  Eq.~(\ref{eq.ap.Phi234}).  The relevant temperature interval, in 
  which $\Phi\sim\epsilon^{-1}$, is not
  accessible in our numerical simulation 
  when $r\neq 0$.  Inset: the same for $q=0.2$ for which the
  transition is discontinuous (notice the different labels in the
  $y$-axis).}
\label{fig.f2f3f4phi}
\end{figure}

In the main panel of Fig.~\ref{fig.f2f3f4phi} we show the behavior of
$\Phi(T)$ for small values of $r$, as described before.  The
simulations are able to correctly detect the long term plateau as long
as it is not too low, otherwise the time to attain it is exceedingly
large and beyond our current computation capabilities. Indeed, the
lowest point for which $\Phi$ was measured in Fig.~\ref{fig.f2f3f4phi}
corresponds to almost $10^{-2}$, that is way before the beginning of
the crossover to a linear behavior. The linear in $\epsilon$ long-term
behavior manifests itself at a much smaller scale, not accessible in
our simulation. On the other hand, in the region where the simulation
was actually performed, the points are indistinguishable from the
$r=0$ case and closer to the $\epsilon^2$ law. Only when $\Phi(t)$
attains a value of the order of $r$, and after an exceedingly long
crossover, the correct $\epsilon$ behavior is observed.  It should be
noticed that larger values of $r$ could diminish these strong
preasymptotic effects, at the expenses of increasing dynamical 
reducibility effects. This effect could be at the origin of the
discrepancy with the value of the measured exponent near the
continuous liquid-glass transition~\cite{MoChKoSc05}.
For the discontinuous liquid-glass transition instead we find that the
square-root anomaly is very well verified.

\subsection{Relaxation times and critical decay exponents}

By measuring $\phi(t)$ at various $T$ and $q$, see
Fig.~\ref{fig.pers_f2f3f4}, one can extract more quantitative
information that will be useful to check
important typical signatures of MCT.
When $q<1/2$, on approaching $\Tg(q)$ from the liquid side, the
persistence $\phi(t)$ presents the characteristic two-step relaxation
form
%
%
%
and we can reasonably estimate the characteristic times as follows:
\begin{equation}
\tau = \int_0^{\infty} \phi(t) dt, \qquad
\tau_{\scriptscriptstyle\beta} = \int_0^{\tc} \phi(t) dt,
\label{eq.deftaubeta}
\end{equation}
where $\tc$ is the time at which persistence  crosses the critical
plateau, $\phi(\tc) = \Phic$. Notice, that according to these
definitions, $\tau$ coincides with $\tau_{\scriptscriptstyle\beta}$ when
$\Phic=0$ (i.e. when the glass transition is continuous). 

\begin{figure}[htbp]
\includegraphics[width=8.5cm]{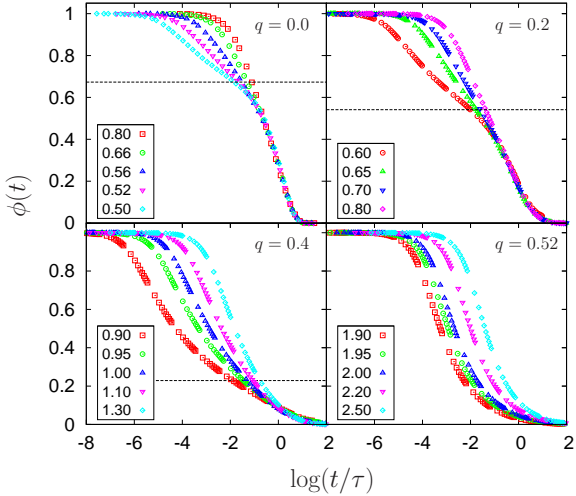}
\caption{Data collapse of persistence data for several values of $q$
  versus the rescaled time, $t/\tau$, where $\tau$ is the
  characteristic time associated with the $\alpha$ relaxation. We
  obtain a good collapse for data such that $\phi(t)<\Phi_c$ and, as
  $\Phi_c\to 0$ when $q\to 1/2$, the collapse region decreases as
  well, disappearing above the dynamical tricritical point. The
  collapsed region is well described by a KWW stretched exponential,
  shown as solid lines for $q=0$ and $0.2$. The stretched exponent is
  $0.5$ for $q=0$ and $0.35$ for $q=0.2$.}
\label{fig.colapso4}
\end{figure}

With these definitions of relaxation times, and after
rescaling time by the structural relaxation time $\tau$,
we can check the existence of universal scaling properties of 
persistence in the late $\alpha$-regime.
For times in the range over which $\phi(t) < \Phic$,
we find a nice data collapse of persistence function once the time is
measured in units of $\tau$, see Fig.~\ref{fig.colapso4}. The range
over which this time-temperature superposition principle holds shrinks
when $q \to 1/2^-$ (and so $\Phic \to 0$), and consequently no simple
scaling form of relaxation data at various temperature can be found in
this limit.  The scaling function is well fitted by a
Kohlrausch-Williams-Watts stretched exponential function with an
exponent that decreases as $q\to 1/2$.




\begin{figure}[htbp]
\includegraphics[width=8.5cm]{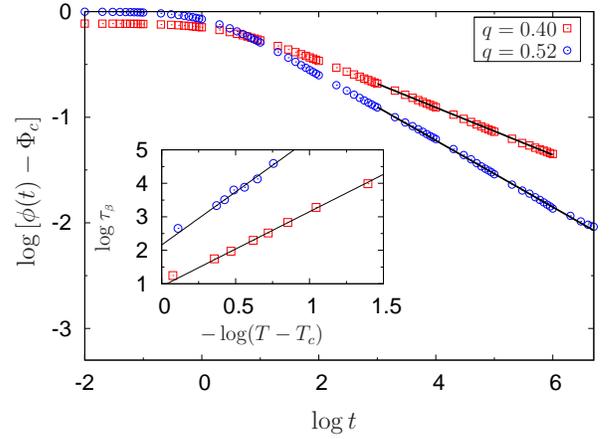}
\caption{Equilibrium relaxation at criticality for discontinuous
  ($q=0.4,\, \Tg\simeq 0.86$) and continuous ($q=0.52,\, \Tg\simeq
  1.72$) ergodic-nonergodic transition: solid lines in the main panels
  are power-law fits with exponents $a \simeq 0.23$ and 0.31,
  respectively. Inset: $\beta$-relaxation time
  $\tau_{\scriptscriptstyle\beta}$ vs temperature difference
  $T-\Tg$. The solid lines are power-law functions with exponents
  $1/2a$ and $1/a$, where the value of $a$ is obtained by the fits in
  the main panel.}
\label{fig.pers_f2f3f4_Tc}
\end{figure}

\begin{figure}[htbp]
\includegraphics[width=8.5cm]{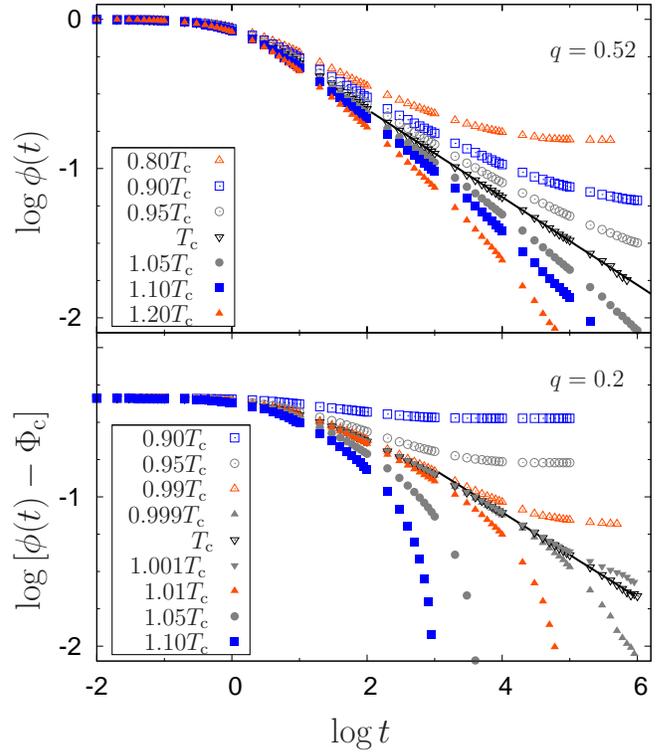}
\caption{Symmetric departure from the critical power law behavior for
  curves at temperatures equidistant from $\Tg$, for $q=0.52$.  The
closer the temperature gets to $\Tg$, the more symmetrical the
curves become and the longer it takes to depart from the critical
power law. An
analogous behavior is observed for $q=0.2$ as well.}
\label{fig.symmetry}
\end{figure}

\begin{figure}[htbp]
\includegraphics[width=8.5cm]{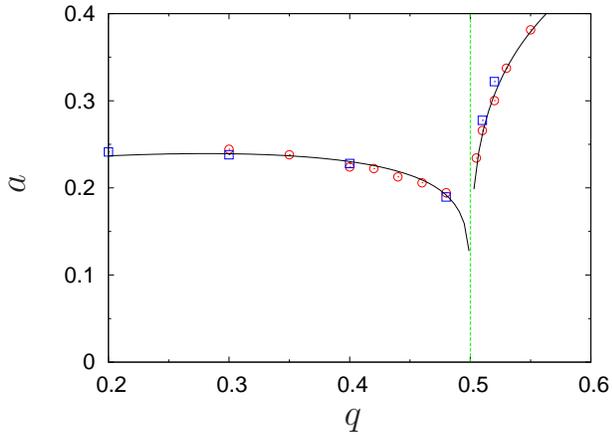}
\caption{Exponent $a$ obtained from the critical relaxation (circles)
  and from the characteristic time as the critical line is approached
  from above (squares).  The solid lines are power-law fits that
  vanish as $q\to 1/2$, consistently with a logarithmic relaxation at that
point.}
\label{fig.a}
\end{figure}

To make a more stringent test of MCT we now examine in detail the
critical dynamics on the glass transition lines, for which MCT
predicts distinctive patterns of universal critical behavior.
When the system is relaxing exactly at the critical temperature
$\Tg(q)$,  MCT predicts a power law decay:
\begin{eqnarray}
\phi(t,\Tg(q))-\Phic(q) \sim t^{-a(q)},
\end{eqnarray}
irrespective of whether $\Phic(q)$ is finite or zero.
The main panel of Fig.~\ref{fig.pers_f2f3f4_Tc} shows that this power
law behavior is very well obeyed, either above and below $q=1/2$.
Slightly above or below $\Tg$, the deviations from
the power law are also predicted by MCT, Fig.~\ref{fig.symmetry}: for 
temperatures equally distant from $\Tg$,
the departure from the critical power law is
symmetric~\cite{Krakoviack07} and starts roughly at the same time.
Accordingly with MCT, the exponent $a(q)$ is intimately related, in a
way that depends on the nature of the glass transition, to the
power-law exponent describing the divergence of the characteristic
time $\tau_{\scriptscriptstyle\beta}$ as the plateau at $\Phic$ is
approached (the so-called $\beta$-relaxation regime): when temperature
gets closer to $\Tg(q)$ one has $\tau_{\scriptscriptstyle\beta} \sim
\epsilon^{-1/2a(q)}$ for the discontinuous transition, and
$\tau_{\scriptscriptstyle\beta} \sim \epsilon^{-1/a(q)}$ for the
continuous one.
These two distinct behaviors are tested in the inset of
Fig.~\ref{fig.pers_f2f3f4_Tc}. Notice that the straight lines in the
inset for describing the data of $\beta$-relaxation time, measured
with Eq.~(\ref{eq.deftaubeta}), are not fitting functions but use the
exponent $a$ of the critical decay law measured in the main panel.
The estimated $\tau$ behaves as a power-law of $\epsilon$
near $\Tg$ and, more importantly, the exponents perfectly agree with
MCT, for both the hybrid and continuous glass transition.
In Fig.~\ref{fig.a} we collect the results of these two independent
ways to measure the exponent $a$, for several values of $q$. They show
excellent agreement. It is also interesting to observe that the
exponent $a$ decreases when $q\to 1/2$, from both side of the
continuous and discontinuous region. Although we were able to fit data
of $a(q)$ with a function that vanishes at $q=1/2$, we were not able
to approach this value of the exponent from the persistence at
$q=1/2$, at least with the sizes and times available to our computer
simulation.

\begin{figure}[htbp]
\includegraphics[width=8.5cm]{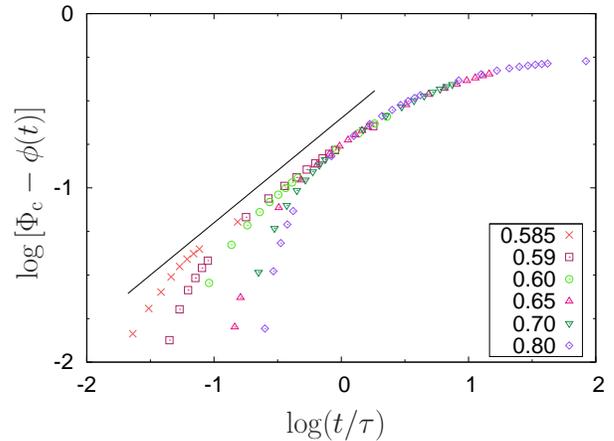}
\caption{Estimate of the exponent $b$ for $q=0.2$ and $r=10^{-3}$. For
  this value of $q$, $T_c\simeq0.58 $ and one obtains $\gamma\simeq
  2.81$, $a\simeq 0.25$ and, using Eq.~(\ref{eq.gamma}), $b\simeq
  0.61$. This predicted value is in very good agreement with the
  simulation, as shown by the solid line whose slope is 0.61.}
\label{fig.b}
\end{figure}

Finally, we test the relation among the various exponents $a$, $b$ and
$\gamma$. For $q=0.2$, the critical temperature is $\Tg\simeq 0.58$,
and we obtain $\gamma\simeq 2.81$ and $a\simeq 0.25$ from the measure
of $\tau$ and $\tau_{\scriptscriptstyle \beta}$, respectively,
Eq.~(\ref{eq.tau_beta}). Thus, Eq.~(\ref{eq.gamma}) gives
$b=(2\gamma-1/a)^{-1}\simeq 0.61$. Plugging this value of $b$ in
Eq.~(\ref{eq.alphabeta_regimes}), we can compare the von Schweidler's
law against the numerical data of simulations, Fig.~\ref{fig.b}.  We
observe that in the late $\beta$ relaxation regime the agreement
between the two is quite good.

\section{Discussion and conclusions}

We have extended the Fredrickson-Andersen facilitation approach to
glassy dynamics by introducing spin mixtures with different types of
facilitation rules~\cite{SeMaCaAr10}.  Depending on the cooperative
nature of the facilitation, they exhibit a crossover from a
discontinuous to a continuous glass transition.  We have carefully
tested by numerical simulation this approach on a Bethe lattice by
analyzing the equilibrium dynamics near the liquid-glass transition
lines.  The results are fully consistent with predictions of MCT and
suggest that the cooperative facilitation approach is the simplest
microscopic on lattice realization of MCT. The existence of such a
correspondence is by no means obvious because MCT is formulated for
many-particle systems obeying Newtonian or Brownian dynamics, while
the facilitation approach concerns particles on lattice obeying an
abstract--though physically motivated--kinetic rule. The close
relation between MCT and systems with quenched
disorder~\cite{KiWo87,KiTh87}, suggests in fact that the dynamical
facilitation description of glassy dynamics is not in conflict with
the RFOT approach. The reason underlying this correspondence (that in
a special case can be exactly proved~\cite{FoKrZa12}) seems to be
quite general~\cite{Sellitto12}, even though it is not trivial to find
the explicit expression of the mapping. To explore further the
correspondence between facilitated and disordered systems it would be
certainly relevant to investigate the possibility of inferring the
exponent $\lambda$, Eq.~(\ref{eq.gamma}), from the MCT kernel or from
the associated ``free energy'' as proposed in
Refs.~\cite{Caltagirone12,PaRi12}.  Also, it is interesting to observe
that, while MCT struggles in the real space high dimensional
(mean-field) limit~\cite{ScSc11,IkMi11}, no such problem is observed
in the Bethe lattice.

The nature of the higher-order glass singularity and other peculiar
dynamical features we observed is primarily controlled by the
underlying bootstrap percolation process describing the formation of a
spanning cluster of frozen spins. The geometric structure of this
cluster can be fractal or compact depending on the nature of the
cooperative dynamics.  The connection with the bootstrap percolation
problem can provide some clues on the fate of dynamical glass transition
in finite spatial dimensions. Although we do not undertake this
delicate task here, we mention that the existence of facilitated
models with a dynamical glass transition in finite
dimensions~\cite{ToBiFi06,JeSc07,OhSa11}, naturally suggests that the
MCT scenario outlined above for higher-order singularity could be
observed also in this circumstance.  Also, we expect that the
dynamical properties near the continuous and discontinuous
liquid-glass transitions are similar to those appearing in the
${\mathsf F}_{13}$ scenario (that differs from the ${\mathsf F}_{12}$ kernel,
Eq.~(\ref{F12}), by the presence of a cubic term instead of
a quadratic one) because the nature of such transitions is
essentially the same in the two schematic models (i.e. a degenerate
and generic ${\mathsf A}_2$ singularity).

Although we focused on Bethe lattice with fixed connectivity and
random distribution of facilitation, it is worth to remark that one
can consider an equivalent variant, in which the Bethe lattice is
diluted and the facilitation is uniform~\cite{SeMaCaAr10}.  For
example, results qualitatively similar to those reported above can be
obtained when $f_i=2$ on every site and the local lattice connectivity
$z_i$ is distributed according to
\begin{eqnarray}
{\cal P}(z_i) &=&(1-q) \ \delta_{z_i,4} + (q-r)\ \delta_{z_i,3} + r\ \delta_{z_i,2}.
\end{eqnarray}
In this case, less connected sites turn out to have a smaller
probability to flip, just as if they were less facilitated.  While the
former variant has the advantage of being more easily implemented in
finite dimensional systems with fixed coordination number the latter
is generally more suitable for systems in which the local mobility
does also depend on the geometrically disordered local environment.
Thus, the crossover between the two glass transitions is obtained by
varying either the local connectivity or the facilitation strength,
and corresponds to a passage from bootstrap to standard percolation
transition.  Whether the fractal cluster of frozen particles can be
assimilated to a gel state is an interesting issue that it would be
worth to address in a future work~\cite{DeFiArCo04,Zaccarelli07}.
Finally, one of the advantages of the present approach is that one can
quantify the contribution given by different types of facilitation to
the glassy features by studying the dynamics of each species of spins
separately. In this respect, it would be interesting to analyze and
compare dynamical heterogeneities close to the continuous and
discontinuous liquid-glass transition.

\bigskip

\acknowledgments We thank D. de Martino and F. Caccioli for their
participation in early stages of this work, and W. G\"otze for
clarifications about MCT. JJA is a member of the INCT-Sistemas
Complexos and is partially supported by the Brazilian agencies Capes,
CNPq and FAPERGS.


\appendix

\section{Calculations for determining the phase diagram and the fraction 
of frozen spins of facilitated spin mixtures}
\label{appendixA}

We report here the detailed calculations to
determine the liquid-glass transitions and the fraction of permanently
frozen spins for models of facilitated ternary mixtures.  With the
notation $\ell+m+n$ we denote a ternary mixture composed by spins with
facilitation $f=\ell,\, m$ and $n$.  For the sake of simplicity, we refer
to a Bethe lattice with branching ratio $k=z-1=3$, where $z$ is the
lattice coordination.  We begin with the facilitated spin system we
focused in the main text and then consider some other, qualitatively
different, representative cases.

\subsection{Ternary mixture $2+3+4$}
\label{ap234}

For the ternary mixture 2+3+4 with facilitation distribution
\begin{equation}
{\cal P}(f_i) = (1-q) \delta_{f_i,2} + (q-r) \delta_{f_i,3} + r
\delta_{f_i,4},
\end{equation}
the probability that a spin is frozen $1-B$ obeys the equation:
\begin{eqnarray}
  1-B &=& p\left[ 1-B^3 - 3(1-q) (1-B) B^2\right].
\label{eq.ap.B234}
\end{eqnarray}
This equation is always satisfied by $B=1$, while an additional
solution with $B<1$ is obtained by solving the quadratic equation
\begin{eqnarray}
  (3q-2) B^2 + B +1-1/p &=& 0,
\label{eq.ap.B234a}
\end{eqnarray}
which gives
\begin{equation}
1-B = \frac{3(1-2q)+\sqrt{1-4(3q-2)(1-1/p) }}{2(2-3q)}.
\end{equation}
The positive sign of the square-root is chosen because $B$ is positive
and should decrease as the temperature goes to zero.  The continuous
transition takes place when the above solution joins the $B=1$
one. This happens when
\begin{equation}
\pc(q) = \frac{1}{3q} .
\end{equation} 
Whereas the vanishing of the square-root argument determines the
discontinuous transition:
\begin{equation}
\pc(q)  = \frac{8-12q}{9-12q}  .
\end{equation} 
Using the relation between the temperature $T$ and the probability $p$
\begin{equation}
  \frac{1}{T} = \ln \frac{p}{1-p},
\end{equation}
one finally gets the dynamical glass transition lines $\Tc(q)$ reported 
in the main text, Eq.~(\ref{eq.pc}).
The crossover between the continuous and discontinuous transition is
obtained when the two expressions of $\Tc(q)$ in Eq.~(\ref{eq.pc})
becomes equal.  This happens at $q=1/2$ and corresponds to the
tricritical temperature, Eq.~(\ref{eq.trc}).

The fraction of permanently frozen spins is
\begin{eqnarray}
  \Phi &=& p\left[ (1-B)^3(4+3B)+6qB^2(1-B)^2+4r (1-B) B^3 \right] \nonumber\\
\label{eq.ap.Phi234}\\
&+&  (1-p)\left[ h^3(4-3h)+6qh^2(1-h)^2+4 r h(1-h)^3
    \right], \nonumber
\end{eqnarray}
where $B$ is given by Eq.~(\ref{eq.ap.B234}) and $h$ is 
\begin{eqnarray}
h = p(1-B)\left[ (1-B)(1-B + 3qB) + 3rB^2  \right].
\end{eqnarray}

\begin{figure}[htbp]
\includegraphics[width=8.5cm]{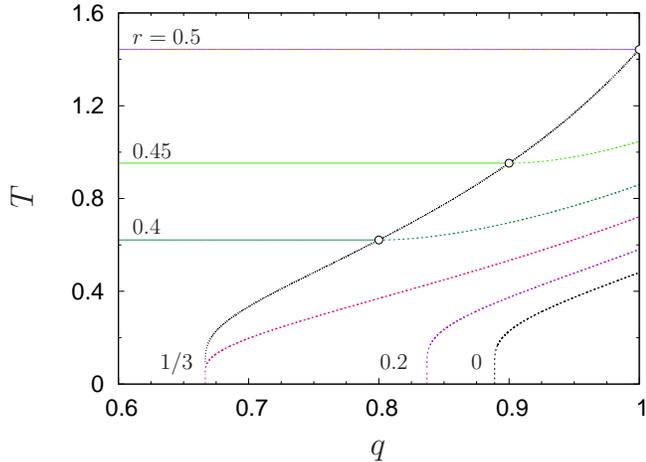}
\caption{Phase diagram, temperature $T$ vs fraction of constrained
  spins $q$, for the ternary mixture 0+2+3 (and 0+2+4) on a Bethe
  lattice with $z=4$ and facilitation as in
  Eq.~(\ref{eq.distr.023}). The dotted/full lines are
  discontinuous/continuous liquid glass transitions for various
  $r$. The line passing through the circles is the tricritical line.}

\label{fig.diag.023}
\end{figure}

\subsection{Ternary mixtures 0+2+3 and 0+2+4}

Next we consider the diluted ternary mixtures of type 0+2+3 and 0+2+4.
These cases are interesting because they provide us with a phase diagram
showing a ``tricritical'' line separating two glass transition {\it
  surfaces}. Let us focus first on the mixture 0+2+3, with 
\begin{equation}
{\cal P}(f_i) = (1-q) \delta_{f_i,0} + (q-r) \delta_{f_i,2} + r
\delta_{f_i,3}.
\label{eq.distr.023}
\end{equation}
The probability $1-B$ that a spin is frozen in this case satisfies the
equation
\begin{equation}
  1-B = p\left[ q(1-B^3) - (q-r) (1-B) 3 B^2\right].
\label{eq.ap.B023}
\end{equation}
Beyond the trivial solution $B=1$ one also finds the extra solution
\begin{equation}
  1-B = \frac{6r-3q+\sqrt{q^2-4(q-1/p)(3r-2q) }}{2(3r-2q)}.
\end{equation}
As mentioned before, the continuous transition occurs when the above
solution joins the $B=1$ one, what happens for $\pc=1/3r$;
while the vanishing of the square-root argument in the latter
determines the discontinuous transition:
\begin{equation}
\pc(q,r) = \frac{4}{3q}\frac{2q-3r}{3q-4r}.
\end{equation}
Using the relation between $p$ and $T$ one finally gets the dynamical 
glass transition surfaces, $\Tc(q,r)$:
\begin{equation}
  \frac{1}{\Tg(q,r)} = \left\{
\begin{array}{ll}
\displaystyle \ln\frac{4(2q-3r)}{12 r(1-q)+q(9q-8)} , & \mbox{\rm
  (discontinuous)}; \\ \\ \displaystyle \ln \frac{1}{3r-1}, &
\mbox{\rm (continuous)}.
\end{array}
\right. 
\label{eq.ap.Tc023}
\end{equation}
Notice that the continuous transition exists in the interval $1/3 \leq
r \leq 2/3$ and does not depend on $q$.  Interestingly, for any $r$ in
the range $1/3 \leq r \leq 1/2$, there is a dynamical tricritical
point whose location in the phase diagram depends on the fraction of
free spins. The location of this tricritical line separating the two
glass transition surfaces is obtained by setting $q=2r$, what gives
\begin{eqnarray}
  \frac{1}{\Ttric(q)} = \ln\frac{2}{3q-2}.
\end{eqnarray}
Sections of the phase diagram are illustrated in the
Fig.~\ref{fig.diag.023}. Since the equation term related to the spins
with $f=3$ is identical to that of spins with $f=4$, the phase diagram
of the mixtures 0+2+3 and 0+2+4 is just the same. While the related
fractions of frozen spins differ: near the continuous transition line
they vanish with a different critical exponent, $\Phi \sim
\epsilon^{\beta}$ with $\beta=2$ for the the mixture 0+2+3 and $\beta
=1$ for the 0+2+4 one. The phase diagram structure for mixtures 1+2+3
and 1+2+4 is qualitatively similar.

\begin{figure}
\includegraphics[width=8.5cm]{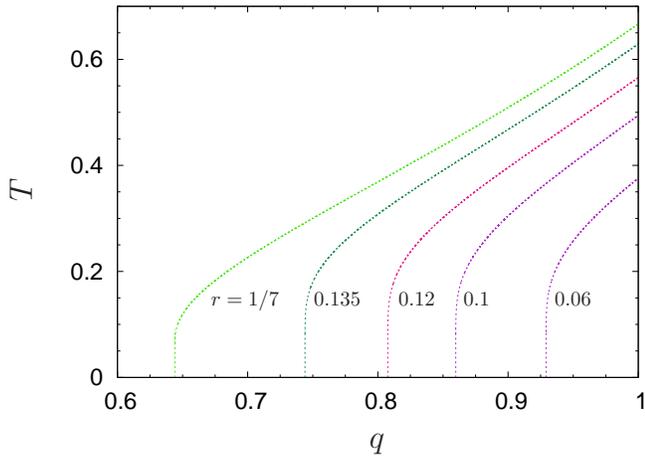}
\caption{Phase diagram for the ternary mixtures 0+1+2 on a Bethe
  lattice with $z=4$ and facilitation as in
  Eq.~(\ref{eq.distr.012}). The glassy region is located below the
  lines which represent discontinuous liquid-glass transitions for
  various $r$.  }
\label{fig.diag.012}
\end{figure}

\begin{figure}
\includegraphics[width=8.5cm]{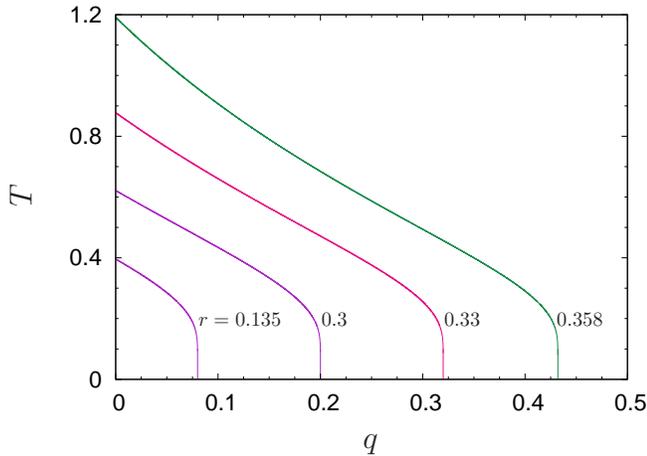}
\caption{Phase diagram for the ternary mixture 0+1+3 (and 0+1+4) on a
  Bethe lattice with $z=4$ and facilitation as in
  Eq.~(\ref{eq.distr.013}). The glassy region is located below the
  lines which represent continuous liquid-glass transitions for
  various $r$.}
\label{fig.diag.013}
\end{figure}

We report here only the fraction of frozen spins for the mixture
0+2+3, which is
\begin{eqnarray}
\Phi &=& pq(1-B)^3( 1+3B) + 6prB^2(1-B)^2 \\ &+& q(1-p)h^3( 4-3h) +
6r(1-p)h^2(1-h)^2, \nonumber
\end{eqnarray}
where $B$ is given by Eq.~(\ref{eq.ap.B023}) and
\begin{equation}
h=p(1-B)^2\left[3rB + q(1-B)\right].
\end{equation}

\subsection{Ternary mixtures 0+1+2, 0+1+3 and 0+1+4}

Finally, for completeness, we consider the ternary mixtures of type
0+1+2, 0+1+3 and 0+1+4.  The calculation proceeds in a manner similar to
those outlined above and, as expected, we find only one type of glass
transition in these mixtures. We report here only the expression of
their critical lines. For the 0+1+2 mixture with facilitation distributed
as
\begin{equation}
{\cal P}(f_i) = (1-q) \delta_{f_i,0} + (q-r) \delta_{f_i,1} + r
\delta_{f_i,2},
\label{eq.distr.012}
\end{equation}
the glass transition is discontinuous and we get
\begin{equation}
  \frac{1}{\Tg(q,r)} = - \ln \left[ r-1 + \frac{(2q-3r)^2}{4q-12r} \right].
\label{eq.ap.Tc012}
\end{equation}
For the 0+1+3 mixture with 
\begin{equation}
{\cal P}(f_i) = (1-q) \delta_{f_i,0} + (q-r) \delta_{f_i,1} + r
\delta_{f_i,3},
\label{eq.distr.013}
\end{equation}
we obtain the continuous glass transition
\begin{equation}
  \frac{1}{\Tg(q,r)} = - \ln (4r-q-1) .
\label{eq.ap.Tc013}
\end{equation}
The above expression is valid for the case 0+1+4 as well.
Sections of the phase diagrams are reported in the
Figs.~\ref{fig.diag.012} and~\ref{fig.diag.013}.  We notice that the
glassy phases generated in the discontinuous and continuous transition
are both pretty stable against dilution (or random damage of the
underlying lattice).  The fraction of free spins can be used in this
type of systems as a way to control the plateau height.  The phase
diagram of binary mixtures can be easily obtained from the previous
calculations by setting one of the weights to zero.
%


\end{document}